\documentclass[twocolumn,showpacs,preprintnumbers,amsmath,amssymb]{revtex4}

\usepackage{graphicx}
\usepackage{dcolumn}
\usepackage{bm}

\begin{document}


\title{Making vortices in dipolar spinor condensates via rapid adiabatic passage}
\author{J.-N. Zhang$^1$, L. He$^1$, H. Pu$^2$, C.-P. Sun$^1$, and S. Yi$^1$}

\affiliation{$^1$Institute of Theoretical Physics, Chinese Academy
of Sciences, Beijing, 100190, China}

\affiliation{$^2$Department of Physics and Astronomy, and Rice Quantum Institute, Rice University, Houston, Texas 77251-189-2, USA}

\begin{abstract}
We propose to the create vortices in spin-1 condensates via magnetic dipole-dipole interaction. Starting with a polarized condensate prepared under large axial magnetic field, we show that by gradually inverting the field, population transfer among different spin states can be realized in a controlled manner. Under optimal condition, we generate a doubly quantized vortex state containing nearly all atoms in the condensate. The resulting vortex state is a direct manifestation of the dipole-dipole interaction and spin textures in spinor condensates. We also point out that the whole process can be qualitatively described by a simple rapid adiabatic passage model.
\end{abstract}

\date{\today}

\maketitle

Ever since the first realization of vortices in an atomic condensate by dynamical phase imprinting~\cite{dynamical}, quantized vortices in quantum gases have attracted great attention. A variety of techniques have been used to generate vortices, including mechanically stirring the atomic cloud with laser beams~\cite{stirring}, rotating asymmetric traps~\cite{rotating}, slicing through the condensate with a fast-moving perturbation~\cite{slicing}, topological phase imprinting~\cite{ketterle}, decay of solitons~\cite{soliton}, coherently transferring orbital angular momentum from photons to atoms~\cite{optical}, and merging multiply trapped condensates~\cite{multi}. None of these schemes, however, relies on the specific forms of atom-atom interactions.

In the present work, we propose a novel scheme to create vortices in a spin-1 condensate by utilizing the magnetic dipole-dipole interaction such that the resulting vortices become a direct manifestation of the underlying dipolar interaction. For convenience, the three spin components of a spin-1 atom are labeled as $\alpha = 0$ and $\pm 1$. Starting with a pure condensate of $\alpha=-1$ atoms prepared under an axial magnetic field along negative $z$-axis, we show that a doubly quantized vortex state in $\alpha=1$ component can be created by gradually inverting the magnetic field from negative to positive. With a careful control of the sweeping rate of the magnetic field and other parameters, the efficiency of atom transfer from $\alpha=-1$ to $1$ component can approach unity. Quite remarkably, as we will show, this dynamical evolution can be understood as a rapid adiabatic passage process described by a simple Landau-Zener tunneling model. Furthermore, due to their very rich physical properties~\cite{yi2,ueda,machida1,yi3,ueda2,santos,bongs}, dipolar spinor condensates have become one of the focuses in the study of quantum gases. In the experiment performed at Stamper-Kurn's group~\cite{kurn}, evidences supporting dipolar effects show up in the form of intriguing spin textures whose detection, however, requires sophisticated imaging techniques. While in our scheme, the vortex state is a much more robust signal and can be readily detected via a straightforward density measurement.


We consider a condensate of $N$ spin-1 $^{87}$Rb atoms. Within the mean-field treatment, the dynamical behavior of the condensate wave functions $\psi_\alpha({\mathbf r})$ is described by (here and henceforth, summation over repeated indices is assumed)
\begin{eqnarray}
i\hbar\frac{\partial\psi_\alpha}{\partial
t}=[T+U+c_0\rho({\mathbf
r})]\psi_\alpha+g\mu_B{\mathbf B}_{\rm eff}({\mathbf r})\cdot{\mathbf
F}_{\alpha\beta}\psi_\beta,\label{gpe}
\end{eqnarray}
where $T=-\hbar^2\nabla^2/(2m)$ with $m$ being the mass of the atom, $U({\mathbf r})=\frac{1}{2}m\omega_0^2(x^2+y^2+\lambda^2z^2)$ is the trapping potential with $\lambda$ being the trap aspect ratio, $\rho({\mathbf r})=\psi_\alpha^*\psi_\alpha$ is the total density of the condensate, and $c_0=4\pi\hbar^2(a_0+2a_2)/(3m)$ characterizes the spin independent collisional interaction with $a_f$ ($f=0,2$) being the $s$-wave scattering length in the combined symmetric channel of total spin $f$. For rubidium atoms, we have $a_0=101.8a_B$ and $a_2=100.4a_B$ with $a_B$ being the Bohr radius. Furthermore, $g(=-1/2)$ is the Land\'{e} g-factor, $\mu_B$ is the Bohr magneton, and ${\mathbf F}$ is the angular momentum operator. The effective field includes the external magnetic field ${\mathbf B}=B(t){\mathbf z}$, and the mean fields originating from the spin-exchange and dipole-dipole interactions
\begin{eqnarray}
{\mathbf B}_{\rm eff}={\mathbf B}+\frac{c_2}{g\mu_B}{\mathbf S}({\mathbf
r})+\frac{c_d}{g\mu_B}\!\!\int\! d{\mathbf r}'\frac{{\mathbf S}({\mathbf
r}')-3\left[{\mathbf S}({\mathbf
r}')\cdot{\mathbf e}\right]{\mathbf e}}{|{\mathbf r}-{\mathbf r}'|^3},\nonumber\\\label{beff}
\end{eqnarray}
where $c_2=4\pi\hbar^2(a_2-a_0)/(3m)$, ${\mathbf S}({\mathbf r})=\psi_\alpha^*{\mathbf F}_{\alpha\beta}\psi_\beta$, ${\mathbf e}=({\mathbf r}-{\mathbf r}')/|{\mathbf r}-{\mathbf r}'|$, and the strength of dipolar interaction is characterized by $c_d=\mu_0\mu_B^2g^2/(4\pi)$ with $\mu_0$ being the vacuum magnetic permeability.

For the numerical results presented in this work, the transverse trapping frequency is taken to be $\omega_0=(2\pi)100\,{\rm Hz}$. We shall focus on three different trap geometries corresponding to prolate, spherical, and oblate trapping potentials with, respectively, $\lambda=0.25$, $1$, and $6$. Unless otherwise stated, the total number of atoms is chosen to be $N=2\times 10^6$.

\begin{figure}
\centering
\includegraphics[width=3.in]{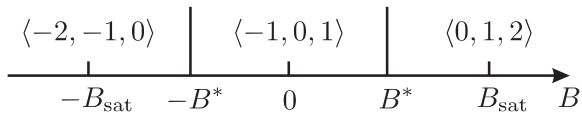}
\caption{Schematic plot of the magnetic field dependence of the ground state structure in an oblate trap.}
\label{phase}
\end{figure}

Let us first recall the ground state structure in magnetic field. For convenience, we assume that the magnetic field is along the negative $z$-axis, i.e., $B<0$. In an oblate trap, as shown in Ref.~\cite{yi3}, the ground state wave function takes the form $\psi_\alpha({\mathbf r})=\sqrt{\rho_\alpha({\mathbf r})}\exp[i(w_\alpha\varphi+\varphi_\alpha)]$, where the densities $\rho_\alpha=|\psi_\alpha|^2$ are axially symmetric, $w_\alpha$ are the winding numbers, $\varphi$ is the azimuthal angle, and $\varphi_\alpha$ are phase angles satisfying $2\varphi_0-\varphi_1-\varphi_{-1}=0$. There exists a critical magnetic field strength $B^*$ such that $\langle w_1,w_0,w_{-1}\rangle=\langle -1,0,1\rangle$ and $\langle -2,-1,0\rangle$ for $0>B>-B^*$ and $B<-B^*$, respectively. Across the critical field $-B^*$, the total energy is continuous, indicating a second-order phase transition. Obviously, for sufficiently strong magnetic field, the system will be polarized. We can define the saturation field strength $B_{\rm sat}>0$ such that when $B<- B_{\rm sat}$, over $99.9\%$ of the population will be in $\alpha =-1$ component. The saturation field is an increasing function of $N$ and $\lambda$. For parameters adopted in this paper, $B_{\rm sat}$ is about several tens of micro-Gauss. Finally, incorporating the results corresponding to $B>0$ case gives us the complete picture of ground state phases in an oblate trap, schematically plotted in Fig.~\ref{phase}.

In a spherical trap, the axial symmetry of the densities $\rho_\alpha$ is lost under weak magnetic field. However, if one increases the magnetic field strength, $\rho_\alpha$ recovers axial symmetry, especially when the system becomes polarized. The saturation field in a spherical trap is only about several micro-Gauss, much lower than that in an oblate trap. Finally, only polarized phase exists in a prolate trap.

\begin{figure}
\centering
\includegraphics[width=3.in]{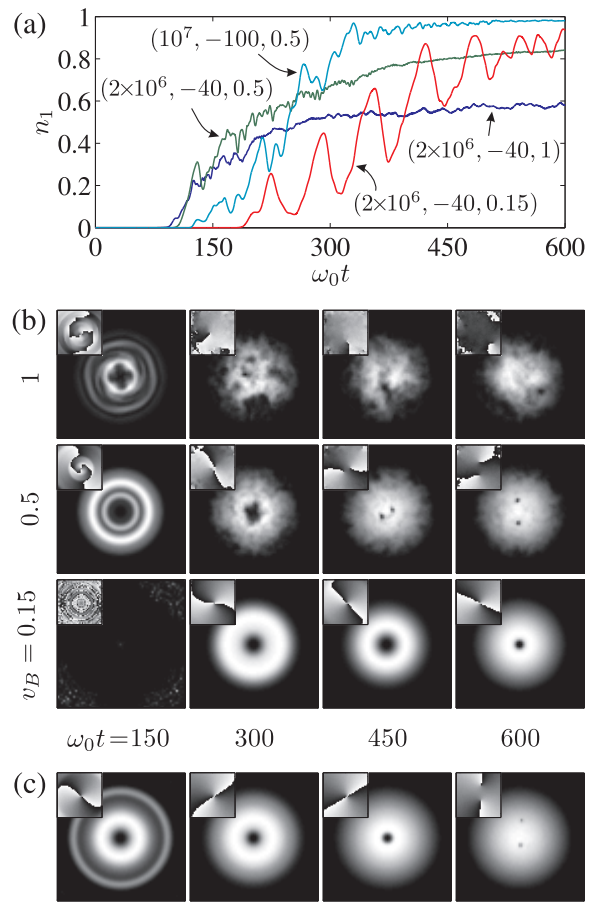}
\caption{(Color online) Typical dynamic behaviors in an oblate trap ($\lambda=6$). (a) The time dependence of reduced atom number $n_{1}$ for various control parameters $(N,B_0,v_B)$. (b) The integrated densities $\bar\rho_1(x,y)$ and phases (insets) of wave function $\psi_{1}(x,y,0)$ for $(N,B_0)=(2\times10^6,-40)$ with various $v_B$'s. (c) Same as (b) except for $N=10^7$, $B_0=-100\,\mu{\rm G}$, and $v_B=0.5$.}
\label{dyn_oblate}
\end{figure}

To study the dynamic properties, we numerically evolve Eq.~(\ref{gpe}) with the initial wave functions being the ground state under field $B_0<-B_{\rm sat}$. The magnetic field is assumed to vary linearly as
\begin{eqnarray}
B(t)=B_0+v_Bt,\nonumber
\end{eqnarray}
where $v_B>0$ (in units of $\mu{\rm G}\cdot\omega_0$) is the sweeping rate of the magnetic field. We shall explore how to control the dynamic behaviors of rubidium condensates by tuning parameters $\lambda$, $N$, $B_0$, and $v_B$.

In Fig. \ref{dyn_oblate} (a), we plot the time dependence of the reduced atom number $n_\alpha=N^{-1}\int d{\mathbf r}\rho_{\alpha}({\mathbf r})$ for $\alpha=1$ spin component in an oblate trap. Since $n_1$ is essentially zero at the beginning, its final value can be regarded as a measure of the atom transfer efficiency. $n_1$ remains negligible until $B$ is close to zero. It then grows with oscillations and eventually reaches some steady-state value. Given $N$ and $B_0$, the asymptotic transfer efficiency increases as one lowers the sweeping rate, and it can be as high as $80\%$ for $N=2\times10^6$ and $v_B=0.15$. However the onset of population transfer occurs earlier for larger $v_{B}$. As we shall show below, the relation between $n_1$ and $v_B$ can be understood using a simple Landau-Zener tunneling (LZT) model of a three-level system.

Figure \ref{dyn_oblate} (b) shows the time dependence of integrated density $$\bar \rho_\alpha(x,y)=\int dz \,\rho_\alpha({\mathbf r}).$$ We see that the density depletion appears at the center of the condensate $\psi_1$. Further examination of the phases of the wave function confirms that it is a doubly quantized vortex which is unstable against splitting into two singly quantized vortices \cite{vortex2}. Indeed one can see that, as the system continues to evolve, the dynamical instability of the wave function $\psi_1$ sets in and the doubly quantized vortex breaks into two vortices. The time that this break happens is sensitive to the magnetic field sweeping rate: It happens at a later time for a lower sweeping rate (i.e., smaller $v_B$).

In Fig.~\ref{dyn_oblate} (c), we show the evolution of the integrated density $\bar{\rho}_1$ for an increased total number of atoms $N=10^7$. In this case, the atom transfer efficiency becomes nearly unity [see Fig.~\ref{dyn_oblate}~(a)] even for a rather large sweeping rate $v_B=0.5$ and a stronger initial magnetic field $B_0=-100\,\mu{\rm G}$. The doubly quantized vortex state survives for a much longer time compared with the previous case with smaller $N$ but the same sweeping rate.

The creation of the doubly quantized vortex can be most easily understood by noting that the dipolar interaction conserves the total angular momentum ${\mathbf J}={\mathbf F}+{\mathbf L}$. When an axial magnetic field is present, $J_z$ is still conserved. For the initial condensate under study, we have $m_J=m_F=-1$. If we assume that all atoms are transferred to $\alpha=1$ state at the end of the process, the spin angular momentum then becomes $m_F=1$. As a result, wave function $\psi_1$ must carry an orbital angular momentum $m_L=-2$, representing a doubly quantized vortex, in order to conserve $J_z$.

\begin{figure}
\centering
\includegraphics[width=2.7in]{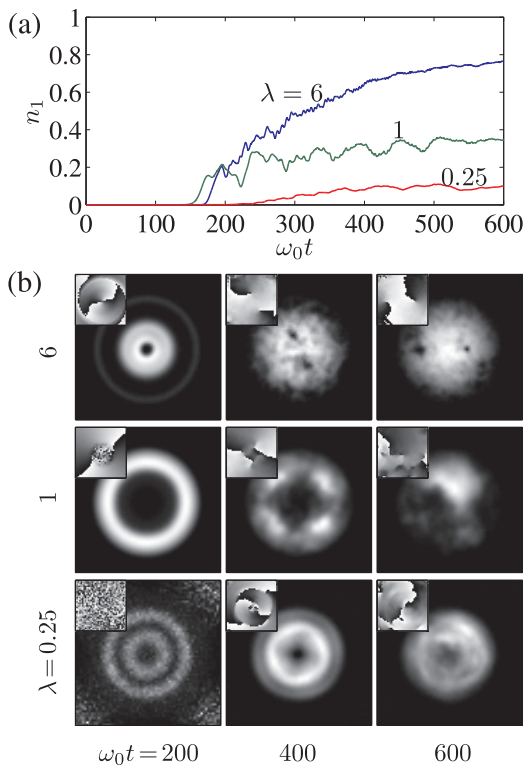}
\caption{(Color online) Time dependence of relative population $n_1$ (a) and integrated density $\bar\rho_1$ (b) in various trapping potentials for $B_0=-60\,\mu{\rm G}$ and $v_B=0.5$. Insets in (b) show the phases of wave function $\psi_1(x,y,0)$.}
\label{traps}
\end{figure}

Next, we turn to study the effects of trap geometry on the population transfer and the wave functions. In Fig.~\ref{traps}, we present the simulation results for all three trap geometries with $B_0=-60\,\mu{\rm G}$ and $v_B=0.5$. Clearly, for given $B_0$ and $v_B$, $n_1$ increases with trap aspect ratio $\lambda$.  From the plots of the density profile in Fig.~\ref{traps} (b), it can be seen that the doubly quantized vortex in a prolate trap is most stable against splitting. In such prolate trap, the population transfer efficiency is low. The vortex core is filled by a large number of spin $\alpha=-1$ atoms, which provides an effective pinning effect that helps to stablize the vortex state \cite{vortexp}. However, in both prolate and spherical traps, the vortices suffer a different kind of instability: The vortex line tends to be distorted and cannot remain straight along the axial direction. A consequence of such instability is to reduce the contrast of the vortex core in the integrated density profile $\bar{\rho}(x,y)$. From these results, we conclude that an oblate trap provides the best candidate to study this phenomenon.

\begin{figure}
\centering
\includegraphics[width=3.in]{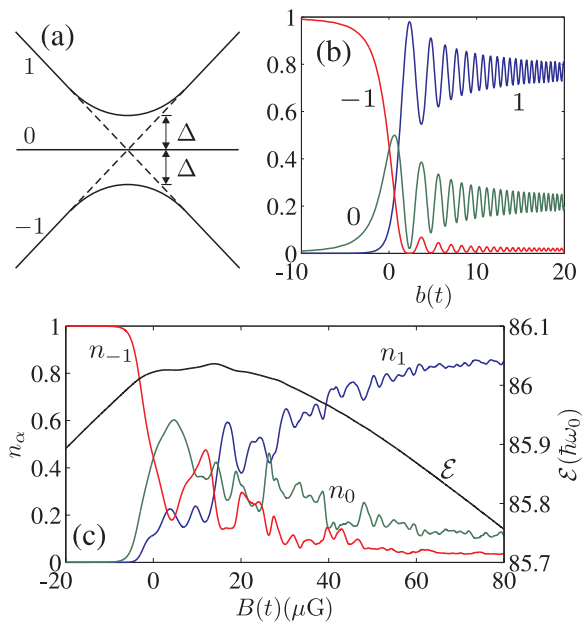}
\caption{(Color online) (a) Eigenenergies of $H_{\rm LZT}$ as a function of axial field $b$. (b) Population dynamics of the LZT model with $b_x=1$, $b_0=-10$, and $v_b=1.5$. (c) Reduced atom numbers $n_\alpha$ and total energy per atom ${\cal E}$ versus magnetic field in an oblate trap ($\lambda=6$). Other parameters are $B_0=-40\,\mu{\rm G}$ and $v_B=0.25$.}
\label{landau}
\end{figure}

To gain more insights into the dynamical properties of the system, let us consider the spin dynamics of a single spin-1 particle under external magnetic field. For the simplest case, it reduces to the model Hamiltonian
\begin{eqnarray}
H_{\rm LZT}=b(t)F_z+b_xF_x \,,\label{model}
\end{eqnarray}
where, without loss of generality, the transverse field $b_x$ is assumed to be a constant and along the $x$-axis. Here all quantities are taken to be dimensionless. The energy spectrum of $H_{\rm LZT}$ is schematically plotted in Fig.~\ref{landau} (a) as a function of the axial magnetic field strength $b$. The presence of $b_x$ changes the three-level crossing at $b=0$ into anti-crossings. The minimum gap between two eigenvalues of $H_{\rm LZT}$ is $\Delta=\sqrt{2}b_x$. Equation (\ref{model}) describes essentially the LZT of three levels~\cite{lze}. Figure~\ref{landau} (b) illustrates an example of the dynamics according to the LZT model. Under a large initial axial field $b_0<0$, the atom is prepared in $\alpha=-1$ state. As we sweep the axial field linearly, $b(t) = b_0 +v_b t$, over the anti-crossings, $\alpha=0$ state is first populated, followed by $\alpha=1$ state. Namely, the wave function bifurcates into the states forming the anti-crossing and thus becomes a coherent superposition of different spin states. This coherent superposition is the cause for the population oscillations which are suppressed at large $t$ limit. The asymptotical transition probability to $\alpha=1$ state is~\cite{lze}
\begin{eqnarray}
p_1=(1-e^{-\pi b_x^2/v_b})^2,\label{lzt}
\end{eqnarray}
an increasing function of $b_x$ while a decreasing one of $v_b$. Moreover, the population oscillations disappear completely for sufficiently small sweeping rate $v_b$ as the adiabatic limit is reached, realizing a rapid adiabatic passage process which represents an important method of transferring population from one quantum state to another.

In Fig. \ref{landau} (c), we present the $B$ dependence of the reduced atom number $n_\alpha$ in a dipolar spinor condensate. Here the transverse field, ${\mathbf B}_{\rm eff}^\perp$, is the transverse component of the effective field. Consequently, it has non-trivial spatial and temporal dependences. Despite the complexity of the condensate system, we still see a remarkable qualitative agreement between the full model and the LZT model: the latter captures all the essential features of the condensate population dynamics. For the parameters used in our full numerical simulations, we always see oscillations in the population dynamics no matter how slowly the magnetic field is swept, meaning we cannot reach the complete adiabatic limit in our full simulation. This is probably due to the accumulated numerical noise. In practice, the lifetime of the condensate (typically on the order of tens of seconds) may also set a constraint on whether adiabatic limit can be reached. Note that in our simulation, we always limit the total time to be about 1 second. In Fig.~\ref{landau} (c), we also plot the total energy per atom ${\cal E}$ as a function of $B$. Again we see that this curve agrees qualitatively with the lowest adiabatic energy level in the LZT model as shown in Fig.~\ref{landau}~(a).

The transition probability Eq. (\ref{lzt}) can also be used to interpret the trap geometry dependence of population transfer efficiency $n_1$. Even through $|c_2|$ is about ten times larger than $c_d$ for rubidium atom, we find in our simulation that the contribution to ${\mathbf B}_{\rm eff}^\perp$ [see Eq.~(\ref{beff})] is mainly provided by the dipole-dipole interaction, and is enhanced by the oblate geometry. Furthermore, the dipolar interaction can also be enhanced by increasing the total number of atoms $N$, which explains the atom number dependence of the transfer efficiency $n_1$.

We remark that in previous studies on the Einstein-de Haas effect of dipolar spinor condensates, the magnetic field is usually inverted suddenly~\cite{ueda2,santos,bongs}. While in this work, the inverting process of the magnetic field is controlled deliberately, such that we may take advantage of the LZT to realize much higher population transfer efficiency and better control of the system.

In conclusion, we have studied the dynamics of a rubidium spinor condensate under a time-dependent axial field. We show that by sweeping the magnetic field and inverting its direction, a doubly quantized vortex containing nearly all atoms in the condensates is created. Despite of the complicated nature of the full system, the population dynamics discussed here can be understood as a rapid adiabatic passage process described by a simple three-level Landau-Zener tunneling model. Since our scheme critically depends on the properties of the dipolar interaction, not only does it provide a new method for generating vortices in atomic condensates, it can also be used as a simple and robust mechanism for the experimental demonstration of dipolar effects in spinor condensates. We hope our work will stimulate experimental efforts along this line. Our future works will also include the study of the controlled spin dynamics of the spin-3 chromium condensate.

We thank Li You for helpful discussion. This work was funded by NSFC (Grants No. 10674141), National 973 program (Grant No. 2006CB921205), and ``Bairen" program of the Chinese Academy of Sciences. HP acknowledges support from NSF and the Robert A. Welch Foundation (Grant No. C-1669).

\end{document}